\documentclass[useAMS,usenatbib,usegraphicx]{mn2e}

\input{psfig.sty}

\title[A relativistic Fe line in PG~1425+267] {Discovery of a
  relativistic Fe line in PG~1425+267 with XMM--Newton and study of
  its short--timescale variability} \author[G.\ Miniutti and A.C. \
Fabian]
{G.~Miniutti\thanks{miniutti@ast.cam.ac.uk} and A.~C.~Fabian \\
  Institute of Astronomy, Madingley Road, Cambridge CB3 0HA}

\pagerange{\pageref{firstpage}--\pageref{lastpage}}
\pubyear{2001}

\usepackage{times}

\begin{document}

\label{firstpage}

 \maketitle

\begin{abstract}
  We report results from the {\it XMM--Newton} observation of the
  radio--loud quasar PG~1425+267 (z=0.366). The EPIC--pn data above
  2~keV exhibit a double--peaked emission feature in the Fe K band.
  The higher energy peak is found at 6.4~keV and is consistent with
  being narrow, while the lower energy one is detected at 5.3~keV and
  is much broader than the detector resolution. We confirm the
  significance of the detection of the broad red part of the line via
  Montecarlo simulations (99.1 per cent confidence level).  We explore
  two possible origins of the line profile i.e. a single relativistic
  iron line from the accretion disc, and the superposition of a narrow
  6.4~keV line from distant material and a relativistic one. We find
  that a contribution from a distant reflector is not required by the
  data. We also perform a time--resolved analysis searching for short
  timescale variability of the emission line. Results tentatively
  suggest that the line is indeed variable on short timescales (at the
  97.3 per cent confidence level according to simulations) and better
  quality data are needed to confirm it on more firm statistical
  grounds. We also detect clear signatures of a warm absorber in the
  soft X--ray energy band.
\end{abstract}


\section{Introduction}

The relativistic broad Fe line is one of the best signatures of the
accretion disc we have been able to collect so far from X--ray
observations and represents a unique tool to probe the strong gravity
regime of General Relativity in a way unaccessible to other
wavelengths. The best examples to date are probably the Seyfert
galaxy MCG--6-30-15 (Tanaka et al 1995; Wilms et al 2001; Fabian et al
2002) and the Galactic black hole candidate XTE~J1650--500 (Miller et
al 2002; Miniutti, Fabian \& Miller 2004) though many other sources of
both classes have broad Fe lines. On the other hand,
{\it XMM--Newton} and Chandra have revealed that a narrow Fe K$\alpha$
emission line is  ubiquitous in the X--ray spectra of bright
Seyfert galaxies (e.g. Page et al 2004; Yaqoob \& Padmanabhan
2004). Such unresolved Fe lines are most likely emitted by a reflector
located at some distance from the central nucleus, i.e. the outermost
accretion disc and/or even more distant matter such as the torus or
the broad line region clouds.

In high luminosity Active Galactic Nuclei such as quasars, the Fe
K$\alpha$ (narrow or broad) was not observed very often before the
advent of a large effective area observatory such as {\it XMM--Newton} (and,
more specifically, the EPIC--pn detector on board of the {\it XMM--Newton}
observatory). In the last few years, the X--ray spectra of a
relatively large sample of quasars (about 40 of them) collected by
{\it XMM--Newton} has revealed that Fe K$\alpha$ emission lines are
relatively common in quasars as well. The recent results by Porquet et
al (2004a) and Jim\'enez--Bail\'on et al (2005) highlight that a Fe
emission line is detected in about half of the quasars, with broad
lines seen in less than 10 per cent of the cases (see also Schartel et
al 2005).  Moreover, the average properties of the Fe lines seen in
quasars seem to be very similar to those inferred for lower luminosity
Seyfert galaxies.

As for broad lines, it should be stressed that the their detection
represents an observational challenge, especially in the most luminous
(distant) quasars due to limited statistics. Broad lines are
relatively easy to detect if the disc is truncated at relatively large
radii and/or the emissivity profile is not much centrally concentrated
so that the line is not very broad and can be distinguished from the
X--ray continuum. On the other hand, if the black hole is rapidly
spinning so that the disc can extend down to small radii, and the
emissivity is centrally concentrated (as it should be), the Fe line
becomes very broad and its little contrast against the continuum
represents a challenge even for large collecting area detectors such
as those on board of {\it XMM--Newton} (see Fabian \& Miniutti 2005
for a discussion).  In this case, the relativistic Fe line can be
detected easily only if the source is characterised by i) super--solar
Fe abundance and/or ii) very large reflection component. It is
interesting, and possibly indicative of a bias induced by our present
observational capabilities, that the best examples we have so far,
even in the case of nearby Seyfert galaxies, always seem to meet the
two above conditions (see e.g.  Fabian et al 2002; 2004).

Here we present results on the discovery of a broad relativistic Fe
line in the X--ray spectrum of a moderate redshift (z=0.366)
radio--loud quasar, PG~1425+267. This quasar is moderately X--ray weak
compared to other radio--loud quasars and exhibits a clear
double--lobed radio structure (Laor et al 1997; Brandt, Laor \& Wills
2000). To our knowledge, the presence of a Fe emission line in
PG~1425+267 was first revealed by ASCA observations (Reeves \& Turner
2000) where an equivalent width of about 120~eV was detected. We also
study the short--timescale variability of the iron emission line in
PG~1425+267.  Short--timescale variability of emission features in the
Fe K band has been reported previously in other active nuclei and
represents one of the most exciting areas of X--ray astronomy, with
the potential of mapping the innermost regions of the accretion flow
with great accuracy (see e.g. Nandra et al 1997; Iwasawa et al 1999;
Turner et al 2002; Petrucci et al 2002; Guainazzi 2003; Yaqoob et al
2003; Longinotti et al 2004; Turner, Kraemer \& Reeves 2004 for some
examples). In one case, such a study revealed evidence for orbital
motion in the relativistic region of the accretion disc close to the
black hole providing a remarkable probe of the dynamics of accreting
matter and of the geometry and nature of the spacetime as predicted by
Einstein's theory of General Relativity (the Seyfert galaxy NGC~3516,
see Iwasawa, Miniutti \& Fabian 2004).

\section{The {\it XMM--Newton} observation}

PG1425+267 was observed by {\it XMM--Newton} during revolution 482 on 2002
July 28 for a total exposure of about 57~ks. The {\it XMM--Newton}
observation is affected by high background throughout the exposure. We
extracted the spectra by selecting good--time intervals in such a way
that the background contribution in the most interesting (see below)
rest--frame 4--7~keV band is less than 4 per cent.  The net exposure
is 28~ks in the EPIC--pn camera, while the observation duration (from
first to last good--time interval) is reduced to about 48~ks. If the
net exposure is reduced by half with a more conservative background
subtraction, the results presented below are unaffected (though error
bars are obviously slightly larger).
Hereafter, errors are quoted at the 90 per cent confidence level for
one parameter of interest unless specified otherwise.

\section{The hard spectrum and Fe K band}

We consider the EPIC--pn spectrum only, because of its higher
sensitivity than MOS in the Fe K band.  We restrict our analysis to
the hard 2--11.6~keV band in the source rest--frame. All fits include
Galactic absorption in the line of sight ($1.76\times
10^{20}$~cm$^{-2}$). We start by considering a single power law model
and find a slope of $\Gamma = 1.46\pm 0.05$.  However, clear residuals
are left below about 5~keV (about 6.8~keV in the source rest--frame).
The data to model ratio for this fit is shown in the top panel of
Fig.~1, where the energy corresponding to neutral iron K$\alpha$
emission (6.4~keV in the rest--frame) is also shown as a vertical
line. The residuals are suggestive
of a asymmetric and possibly double--peaked iron line profile such as
that expected from reflection in an accretion disc.

\begin{figure}
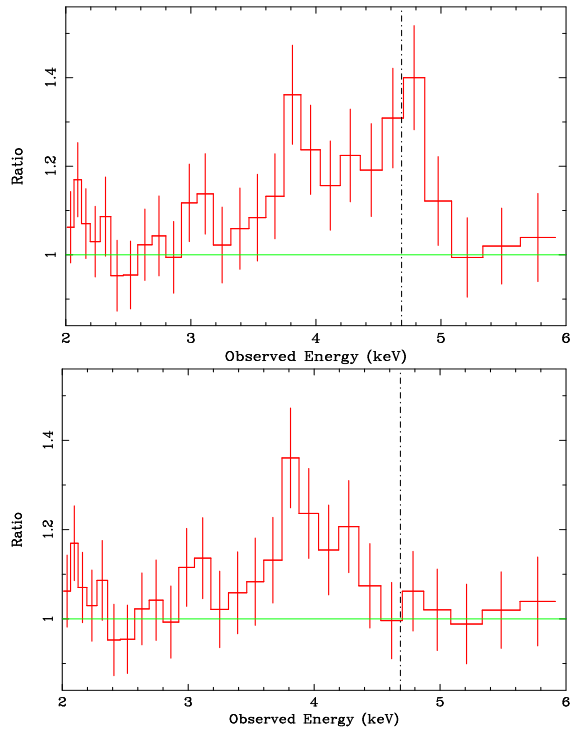

\begin{center}
 \includegraphics[width=0.27\textwidth,height=0.42\textwidth,angle=-90]{Ratio_1.cps}
 \includegraphics[width=0.27\textwidth,height=0.42\textwidth,angle=-90]{Ratio_2.cps}
 \caption{In the top panel, we show the data to model ratio for
   a model comprising  Galactic absorption plus power law. Clear
   residuals in the form of a double--peaked emission feature are
   seen in the 3.4-5.1~keV band (4.6--7~keV~keV in the
   rest--frame). In the bottom panel the ratio is relative to a model
   including one Gaussian emission line at $\sim$~6.4~keV. The
   vertical line shows the rest--frame energy of the neutral Fe
   K$\alpha$ emission line (6.4~keV). Data have been rebinned for
   visual clarity.}
\end{center}
\end{figure}

We then add a Gaussian emission line to the simple continuum model. We
obtain an improvement of $\Delta\chi^2 = 13$ for 3 degrees of freedom
(dof) for a line energy of $6.41\pm 0.17$~keV. Throughout the paper,
line energies are given in the rest--frame unless specified otherwise.
The line is resolved and has a width of
$\sigma=0.28^{+0.41}_{-0.17}$~keV. We measure a line equivalent width
of about 150~eV. However, this value has to be corrected because of
the galaxy redshift and is about 205~eV at 6.4~keV in the rest--frame.
Residuals are still present in the form of a broad emission feature at
lower energy, as shown in the bottom panel of Fig.~1. We then add a
second Gaussian emission line to account for the lower energy
residuals. This improves the statistics by a further $\Delta\chi^2=11$
for 3 more free parameters with a final result of $\chi^2 =281$ for
304 dof. The second Gaussian line is redshifted with respect to
6.4~keV with energy of $5.3\pm 0.2$~keV and is broad
($\sigma=0.4^{+0.7}_{-0.2}$~keV) with respect to the EPIC--pn spectral
resolution. The equivalent width of this lower energy line is about
180~eV in the source rest--frame. Now that both lines are included, we
re--compute best--fitting parameters and errors and give our final
results in Table~1. With the addition of the lower energy broad line,
the 6.4~keV emission line becomes consistent with being unresolved
(less than 400~eV in width).

We also tested a different continuum model in which the primary
continuum is partially covered by a column of gas local to the source.
This model is implemented by adding the {\small{{ZPCFABS}}} model in
{\small{{XSPEC}}} to the power law continuum plus Galactic absorption.
The partial covering model has column density ($N_H$) and covering
fraction ($f_c$) of the absorber as free parameters and has the
potential of reducing the strength of the broad redshifted emission
line (because it induces curvature in the continuum). We also include
a $\sim$~6.4~keV Gaussian emission line. We obtain an acceptable fit
($\chi^2=286$ for 305 dof to be compared with $\chi^2=281$ for 304 dof
for the double Gaussian model). The partial covering parameters are
not well constrained with an upper limit on the column density of $N_H
< 2 \times 10^{23}$~cm$^{-2}$ and a covering fraction between 30 and
95 per cent. Moreover, residuals are still present around 5.3~keV.
This is because, the emission at 5.3~keV (see bottom panel of Fig.~1)
is more peaked than predicted by a partial coverer. Indeed, the
addition of a second Gaussian emission line at 5.3~keV marginally
improves the partial covering fit by $\Delta\chi^2=6$ for 3 more free
parameters. When the second Gaussian is added, the partial covering
parameters become totally unconstrained with best--fitting values
tending to eliminate any absorber contribution. Therefore, partial
covering potentially reduces the significance of the 5.3~keV line but,
though not excluded, such a scenario is certainly not required by the
data which do prefer a solution in terms of the double Gaussian model.
Better quality data are needed to clarify this issue with less
ambiguity. In the analysis below we shall consider a simpler continuum
model in which the X--ray continuum does not suffer from partial
covering complex absorption.

The overall line profile can be better illustrated as follows: we fit
the data with the power law continuum model (plus cold absorption with
column fixed to the Galactic value), add a Gaussian emission line
model, and step on line energy and normalisation in the ranges
$4$--$8$~keV and $0$--$2.5\times 10^{-5}$~ph~cm$^{-2}$~s$^{-1}$
respectively\footnote{We fix the width of the Gaussian filter to a
  value compatible with both detected lines (200~eV, see Table~1).}. We
then record the $\Delta\chi^2$ with respect to the original continuum
model and compute the confidence contour levels. In Fig.~2, we show
the result of this exercise. The contours represent an improvement in
$\chi^2$ by 4.61, 6.17, 9.21, and 11.8 for two more degrees of freedom
(from the outermost to the innermost) and the vertical lines are the
best--fitting energies from the double Gaussian model.

\subsection{Statistical significance of the redshifted broad line}

A narrow iron line at 6.4~keV is ubiquitous in the X--ray spectra of
AGNs, and begins to be detected often in quasars as well (Page et al
2004; Yaqoob \& Padmanabhan 2004; Jim\'enez--Bail\'on et al 2005). In
our double Gaussian best--fit model, the width of the 6.4~keV line is
unconstrained, and only an upper limit of 400~eV can be computed.  The
line energy and width are thus consistent with that of neutral Fe
K$\alpha$ emission from distant material (torus or broad line region).

\begin{figure}
\begin{center}
 \includegraphics[width=0.27\textwidth,height=0.42\textwidth,angle=-90]{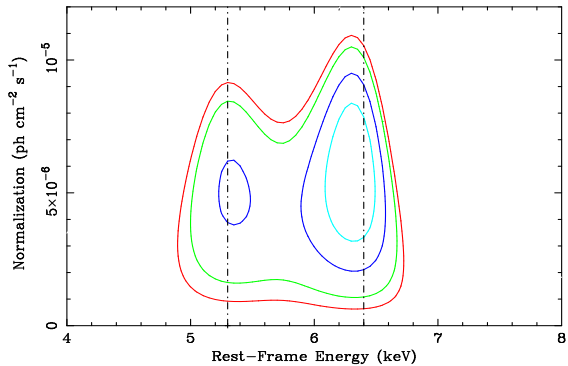}
 \caption{Contours in the flux--energy plane when a Gaussian filter is
   applied to the data modelled with a power law continuum. The
   contours represent an improvement in $\chi^2$ of 4.61, 6.17, 9.21,
   11.8 for the addition of two parameters. The vertical lines are the
   best--fit energies of the emission lines when a phenomenological
   double Gaussian model is used.}
\end{center}
\end{figure}

However, the overall line profile (see top panel of Fig.~1 and Fig.~2)
is reminiscent of a double--peaked line profile such as that
theoretically expected from the accretion disc and shaped by Doppler
and gravitational energy shifts due to high velocity and strong
gravity in the vicinity of the central black hole. The detection of a
relativistic Fe line from the accretion disc in a radio--loud quasar
is much more interesting than that of a narrow Fe line and therefore
deserves a detailed significance study. Below, we perform a test on
the significance of the broad part of the line profile making use of
Montecarlo simulations.

\subsubsection{Montecarlo simulations}

We carried out a test on the broad line detection significance by
extending the method presented by Porquet et al (2004). The
pessimistic assumption is that the blue peak at 6.4~keV is completely
due to unresolved narrow emission (e.g. reflection from
distant matter) and is not part of a relativistic profile.  We here
test only the significance of the broad part of the line (that
modelled by the 5.3~keV Gaussian line in the double Gaussian model).
If the overall line profile is due to a relativistic double--peaked
line without much contribution from an unresolved component, the test
clearly underestimates the significance of the relativistic line and
represent a most conservative lower limit.

We use the best--fitting parameters of a power law plus narrow 6.4~keV
model as our null hypothesis. The 6.4~keV peak of the line is well
reproduced by the model. As throughout the paper, the model includes
Galactic absorption with fixed column density. As mentioned, residuals
do appear as a broad emission feature at 5.3~keV (bottom panel of
Fig.~1) and the purpose of the following exercise is to assess the
probability that they are due to noise in the data. The procedure is
as follows:

\begin{enumerate}
\item starting from the null hypothesis model, we simulate a spectrum
  with the same exposure as in the data and subtract the appropriate
  background;
\item the spectrum is fitted with the model used to generate it, and
  parameters are recorded, providing a new and refined null hypothesis
  model which differs from the original due to photon statistics only;
\item we use this refined model to generate a simulated spectrum,
  subtract the appropriate background, fit it with the power law plus
  narrow 6.4~keV line, and record the $\chi^2$. We then add a second
  Gaussian emission line, vary its energy in the 4--8~keV range, fit
  the data at each step with line width and normalisation free to vary, and we
  record the best--fitting parameters and the maximum $\Delta\chi^2$;
\item we repeat the above steps 1000 times and obtain the distribution
  of $\Delta\chi^2$, to be compared with the result obtained from the
  data. The addition of the second Gaussian in the data gives an
  improvement of $\Delta\chi^2 =11.4$ with respect to the null
  hypothesis model. 
\end{enumerate}
This procedure gives us an estimate on the significance of the broad
red wing of the line under the hypothesis that a narrow 6.4~keV line
is present. Moreover, step (ii) ensures that the uncertainty
in the null hypothesis is properly taken into account because each
spectrum used to compare with the data is obtained from a different
realisation of the hypothesis, which differs from the original due to
photon counting statistics.

\begin{figure}
\begin{center}
 \includegraphics[width=0.27\textwidth,height=0.42\textwidth,angle=-90]{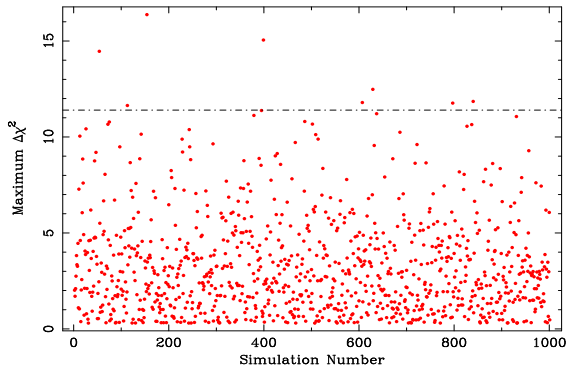}
 \caption{Distribution of the maximum $\Delta\chi^2$ obtained when the
   1000 simulated spectra are fitted with the null hypothesis model
   plus an additional Gaussian emission line. The $\Delta\chi^2$ is
   with respect to the null hypothesis fit. The horizontal line is the
   result obtained by applying the same procedure to the data. Only 9
   spectra show evidence for an additional emission line with the same
   or higher $\Delta\chi^2$ as that obtained in the data. Therefore,
   the broad part of the line seen in the real data is significant at
   the 99.1 per cent level. }
\end{center}
\end{figure}

In Fig.~3 we show the measured $\Delta\chi^2$ distribution from the
1000 simulated spectra and compare it with the result from the data
(horizontal line).  The simulations show that the probability to
detect a line (broad or narrow) in the whole 4--8~keV band and
producing the same (or better) statistics than the broad part of the
line in the real data is 9/1000, i.e. the broad part of the line is
significant at the 99.1 per cent level. As mentioned, this figure
represents the most conservative lower limit for the significance of
the relativistic line as a whole.

\subsection{Direct spectral fitting}

The observed line profile (see Fig.~1 and 2) is very similar to that
expected from an accretion disc extending down in the relativistic
region close to the black hole. As demonstrated above, the broad part
of the line is significant at the 99.1 per cent level.  The 6.4~keV
peak is however consistent with being unresolved. We then test two
different possible scenarios by direct spectral fitting. In the first
one, the whole line profile is due to reflection from the accretion
disc, while in the second case, we allow a narrow 6.4~keV component to
be present.

\subsubsection{A single relativistic iron line}

\begin{table}
\begin{center}
\begin{tabular}{lcc}
\hline
\hline
 Parameter & Blue Line & Red Line \\
\hline
& TWO GAUSSIAN& \\
\hline
$E$ (keV) & $6.4\pm 0.2$ & $5.3 \pm 0.2$ \\
$\sigma$ (eV) & $<400$ & $400^{+700}_{-200}$ \\
EW (eV) & $205^{+135}_{-150}$ &$180^{+280}_{-110}$\\
 $\chi^2/{\rm{dof}}=281/304$  & & \\
\hline
& DISKLINE & \\
\hline
$E$ (keV) & -- & $6.4^f$\\
$\sigma$ (eV) & -- & -- \\
EW (eV) & -- & $410^{+240}_{-150}$ \\
$r_{\rm{in}}$ ($r_g$) & -- & $6^f$ \\
$r_{\rm{out}}$ ($r_g$) & -- & $1000^f$ \\
$q$  & -- & $4.1^{+2.5}_{-1.4}$ \\
$i$ (degrees) & -- & $34^{+5}_{-9}$ \\
$\chi^2 /{\rm{dof}}=282/307$ & & \\
\hline
& GAUSSIAN + DISKLINE & \\
\hline
$E$ (keV) & $6.6 ^{+0.37p}_{-0.2p}$& $6.4^f$\\
$\sigma$ (eV) & $10^f$& -- \\
EW (eV) & $<80$& $400^{+300}_{-180}$\\
$r_{\rm{in}}$ ($r_g$) & -- & $6^f$\\
$r_{\rm{out}}$ ($r_g$) & -- & $1000^f$\\
$q$  & -- & $> 2.7$ \\
$i$ (degrees) & -- & $<38$ \\
 $\chi^2/{\rm{dof}} = 282/305$ & & \\
\hline
\hline
\end{tabular}
\caption{Results of the spectral fitting of the hard EPIC--pn spectrum
  of PG~1425+267 with the different models we explored. Here the
  symbol $^f$ means that the parameter has been fixed, while the
  symbol $p$ denotes that the parameter pegged to the lower and/or
  higher allowed boundaries. Energies and equivalent widths (EW) are
  corrected for the galaxy redshift.}
\end{center}
\end{table}

The single relativistic line hypothesis can be tested by re--fitting
the data with a {{\small{DISKLINE}}} component only (Fabian et al
1989). The model describes the line profile expected if the Fe line is
emitted from the accretion disc around a non--rotating black hole. We
keep the energy of the line fixed at 6.4~keV and fix the outer and
inner disc radii to $1000~r_g$ and $6~r_g$ respectively (as
appropriate if the disc extends down to the innermost stable circular
orbit). The results of this test are reported in Table~1 and the fit
is of the same statistical quality than that with two Gaussian emission
lines (which is however more phenomenological than physically
motivated). The F--test for the addition of the relativistic line with
respect to the power law continuum gives a significance larger than
99.99 per cent.  The disc inclination is well constrained
($i=34^{+5}_{-9}$ degrees) and the emissivity profile is only
marginally steeper than $q=2.5$.

We also tested if emission from radii smaller than $6~r_g$ is required
by using the {{\small{LAOR}}} model, appropriate for Kerr black
holes (Laor 1991). If so, this would suggest that the black hole is
spinning because the accretion disc can extend within $6~r_g$ only if
the spin is different from zero. However, we did not find any
conclusive result, with a 90 per cent upper limit on the inner disc
radius of $16~r_g$ (the outer disc radius is fixed at its maximum
allowed value of $400~r_g$). If the emissivity index is fixed at its
standard value of $q=3$ the upper limit reduces to $8.5~r_g$ which is
still consistent with a disc around a non--spinning black hole. Thus,
we conclude that the data are consistent with any black hole spin from
zero to maximal, and that black hole spin is not required by the data.

\subsubsection{A narrow and a  relativistic
  broad iron line}

To explore the possibility that the 6.4~keV peak of the line is due
(or partly due) to the presence of an unresolved narrow component
(e.g. from distant matter), we consider a composite model with a
narrow Gaussian emission line and a relativistic one. We fix the width
of the Gaussian line to 10~eV and retain the {{\small{DISKLINE}}}
model as above.  We find a good fit with the same statistical quality
as the previous one. However, many of the parameters are unconstrained
because of a clear degeneracy. In particular, since the disc
inclination is related to the blue horn of the relativistic line, it
can not be constrained simultaneously with the narrow Gaussian energy
and flux. The same is true for the emissivity index. Therefore the
narrow Gaussian energy ranges in all the allowed range
(6.4--6.97~keV), its normalisation is only an upper limit, and both
the disc emissivity and inclination are measured only as a lower and
upper limit respectively. We therefore conclude that some contribution
to the blue peak from a narrow Gaussian emission line can not be
excluded but is not required by the data.

\subsubsection{The reflection continuum}

The most likely interpretation is that we are observing Fe emission
whose detailed line profile is shaped by special and general
relativistic effects in the inner regions of the accretion disc. Since
the emission line we observe is a signature of X--ray reflection, we
replace the power law continuum with a {{\small{PEXRAV}}}
reflection model (Magdziarz \& Zdziarski 1995) with cut--off energy
fixed to 100~keV and solar abundances. If the reflector is the
accretion disc, as it seems likely from the line profile, the
reflection continuum is also broadened and blurred by the relativistic
effects. We then apply self--consistently the same relativistic
blurring to the line and to the reflection continuum and show in
Fig.~4 the confidence contour for photon index and reflection
fraction. A reflection continuum is not required by the data, but we
cannot exclude it, with a 90 per cent upper limit on the reflection
fraction of $\sim$~1.2. Notice that the inclusion of the reflection
continuum does not change the parameters of the relativistic Fe line.
With our best--fit {{\small{DISKLINE}}} model, the
(observed--frame) 2--10~keV flux turns out to be $1.6 \times
10^{-12}$~erg~cm$^{-2}$~s$^{-1}$, corresponding to a rest--frame
luminosity of $6.4 \times 10^{44}$~erg~s$^{-1}$.

\begin{figure}
\begin{center}
 \includegraphics[width=0.27\textwidth,height=0.42\textwidth,angle=-90]{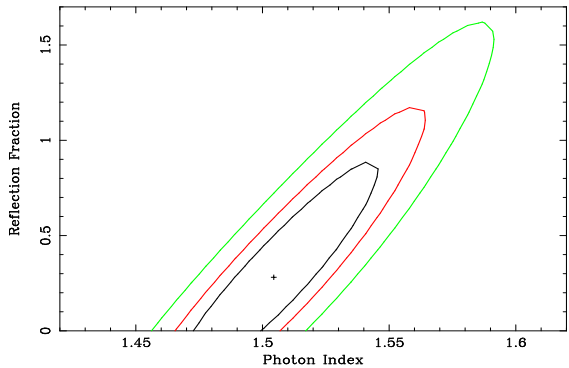}
 \caption{Confidence contour for the (relativistically blurred)
   reflection fraction and photon index. The contours represent region
 of 68, 90, and 99 per cent confidence.}
\end{center}
\end{figure}

\section{Short--timescale variability of the line profile?}

Time--resolved spectroscopy has the great potential of revealing more
clearly the origin of the emission feature. If the Fe line we see in
the data is coming from the inner accretion disc, some variability
could be present. We then investigate below such a possibility by
splitting the observation in different equal duration intervals.

\subsection{Splitting the observation in two halves}

We split the observation in two equal duration intervals about 24~ks
long and subtract from the two spectra the corresponding background.
Due to background flares, the net exposure in the two spectra is
slightly different and much shorter than 24~ks. We fit the two spectra
with the same power law model plus Galactic absorption as above and
add Gaussian emission lines if required.  We also repeat the procedure
used to produce to Fig.~2 and present our results in Fig.~5. We find
one significant emission line in the first half of the observation at
$6.5 \pm 0.2$~keV, while a lower energy line at $5.4\pm 0.3$~keV is
detected only marginally. In the second half, two emission lines are
clearly detected at $5.1\pm 0.2$~keV, and $6.2\pm 0.3$~keV. The
different styles of the vertical lines distinguish between clear and
marginal detections.

The line profile appears to be different in the two halves of the
observation with the strongest of the peaks seen at $6.5 \pm 0.2$~keV
in the first half and at $5.1^{+0.2}_{-0.1}$~keV in the second.
However, it is clear that claiming variability on firm statistical
grounds is a difficult task because emission above the continuum seems
to be present at any given energy between 4.8~keV and 6.6~keV in both
intervals. In fact the intensity of the stronger line at 5.1~keV in
the second half is $8.4^{+6.2}_{-5.4} \times
10^{-6}$~ph~cm$^{-2}$~s$^{-1}$. If a line with the same energy (and
width) is fitted to the first half spectrum, we obtain a 90 per cent
upper limit of $4.5\times 10^{-6}$~ph~cm$^{-2}$~s$^{-1}$ on its
intensity. The two intensities are (though somewhat marginally)
consistent with each other, so that no variability can be claimed. A
detailed discussion on the significance of the variability is deferred
to a shorter timescale analysis below.

\begin{figure}
\begin{center}
 \includegraphics[width=0.26\textwidth,height=0.42\textwidth,angle=-90]{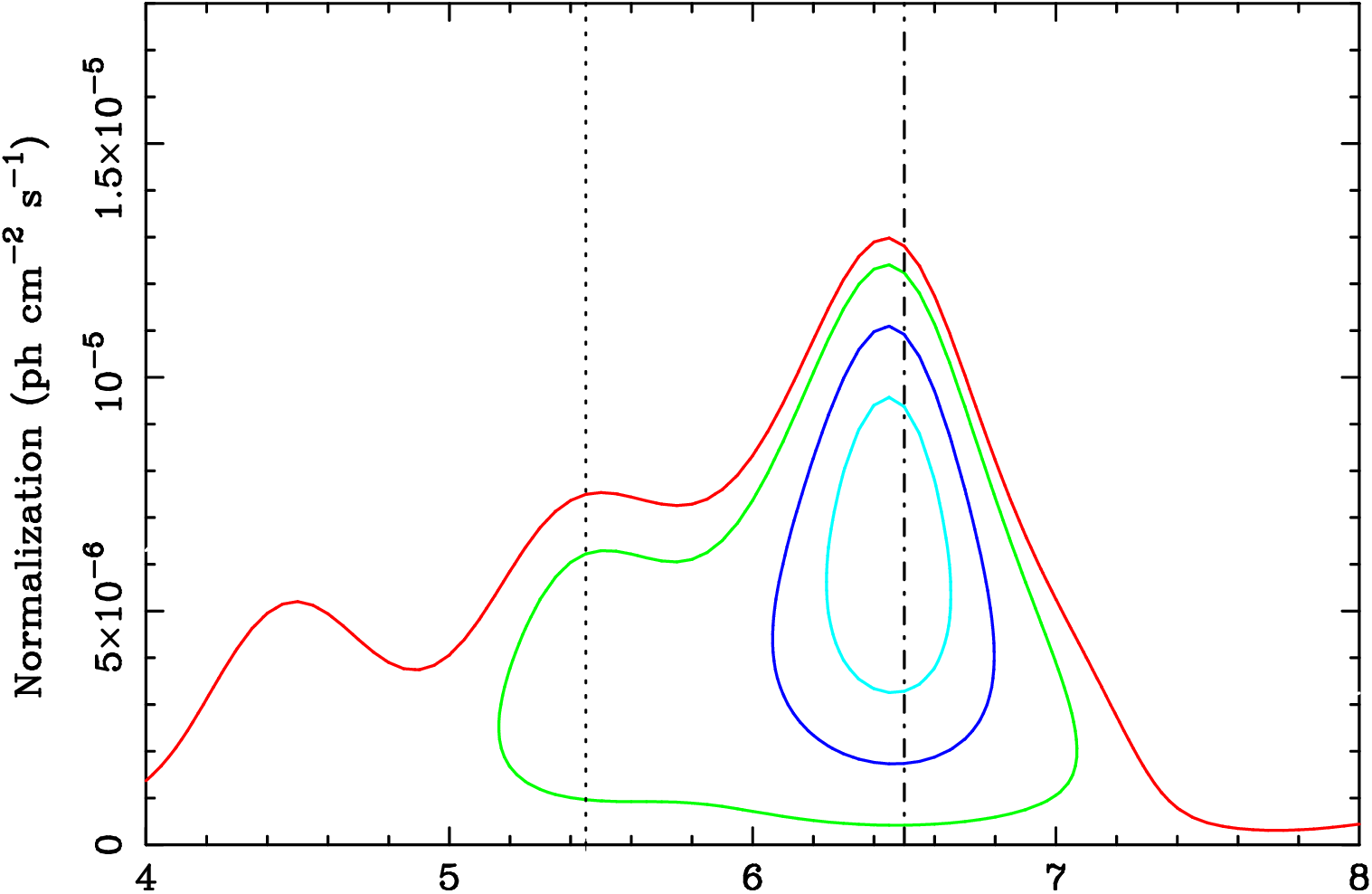}
 \includegraphics[width=0.27\textwidth,height=0.42\textwidth,angle=-90]{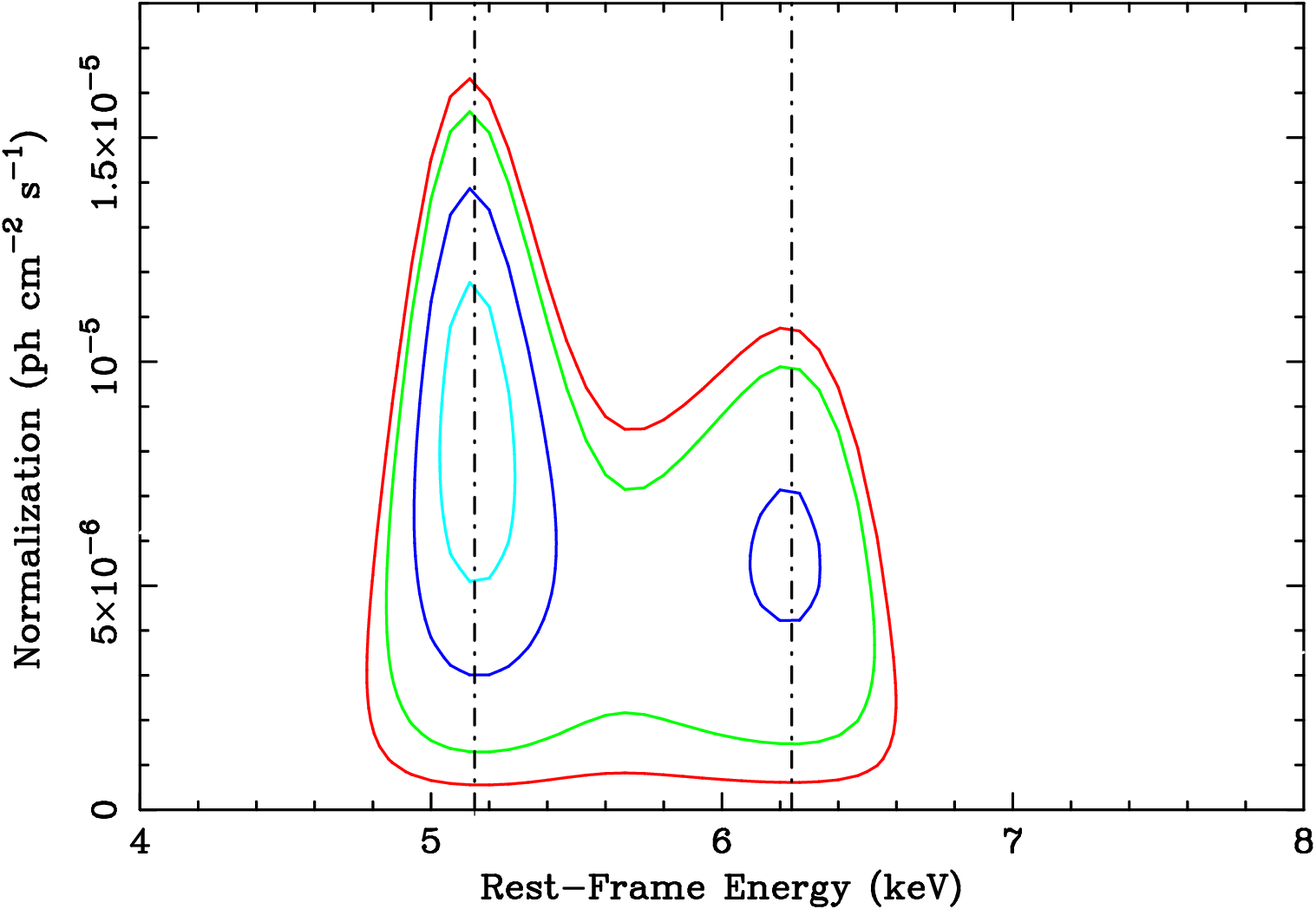}
 \caption{Same as Fig.~2 but for the first half (top) and the second
   half (bottom) of the observation. The vertical lines are the
   best--fit energies from the spectral fits with a power law plus
   Gaussian(s) model. Different line styles are attributed to secure
   and marginal detections.}
\end{center}
\end{figure}

\begin{figure*}
\begin{center}
\centerline{\mbox{ 
\psfig{figure=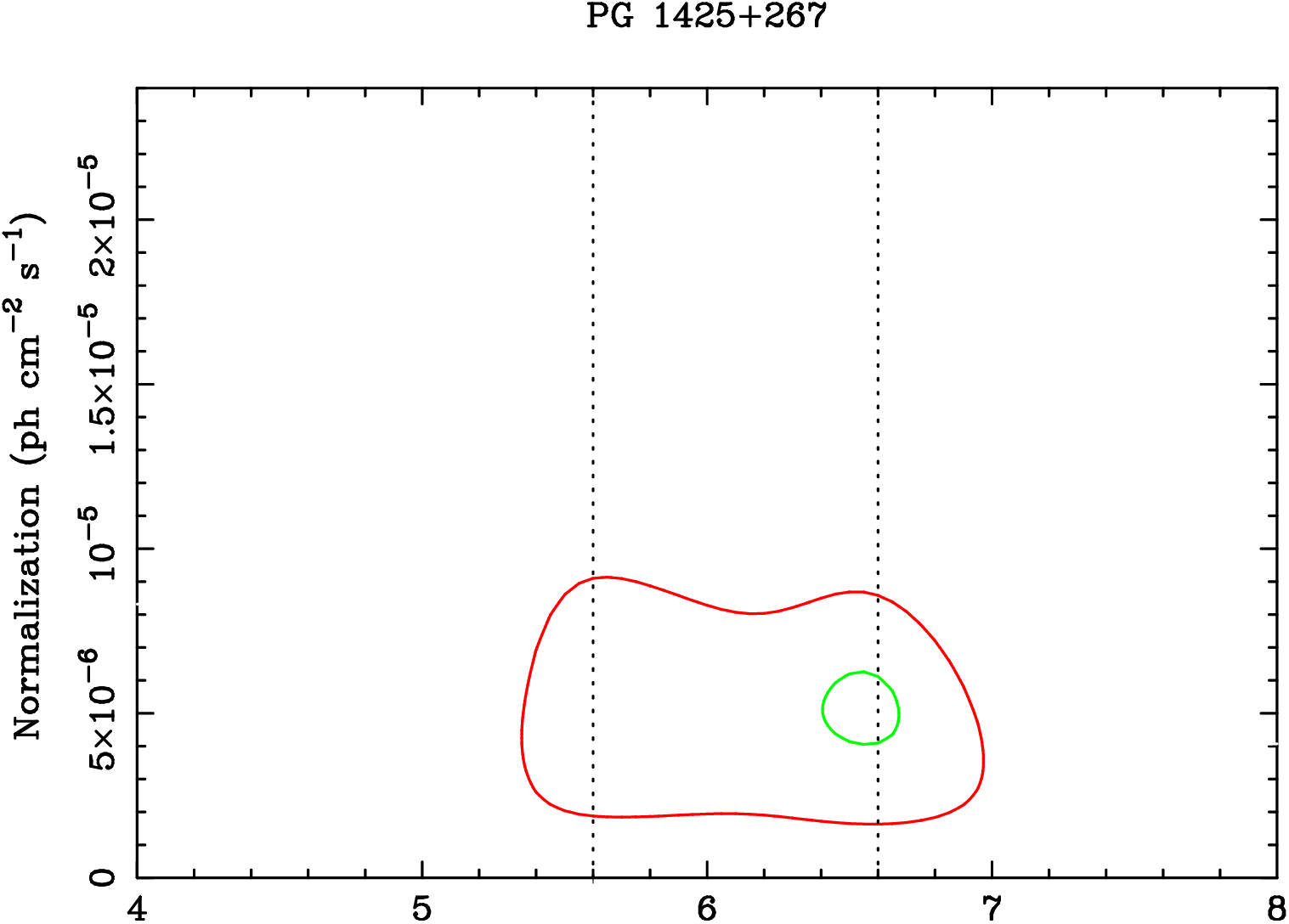,width=0.42\textwidth,height=0.28\textwidth,angle=-90}
\hspace{1.0cm}
\psfig{figure=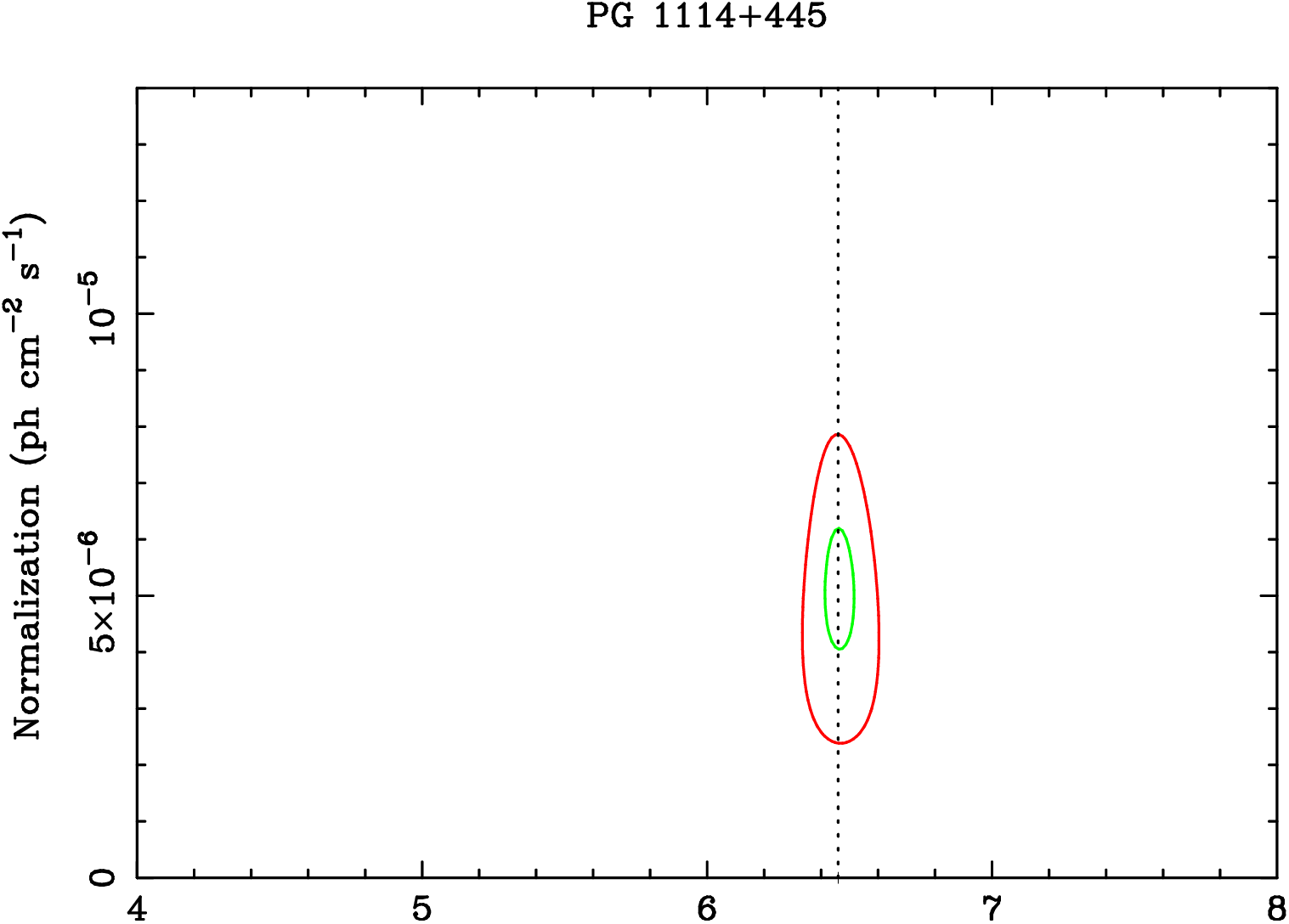,width=0.42\textwidth,height=0.28\textwidth,angle=-90} 
}}
\centerline{\mbox{ 
\psfig{figure=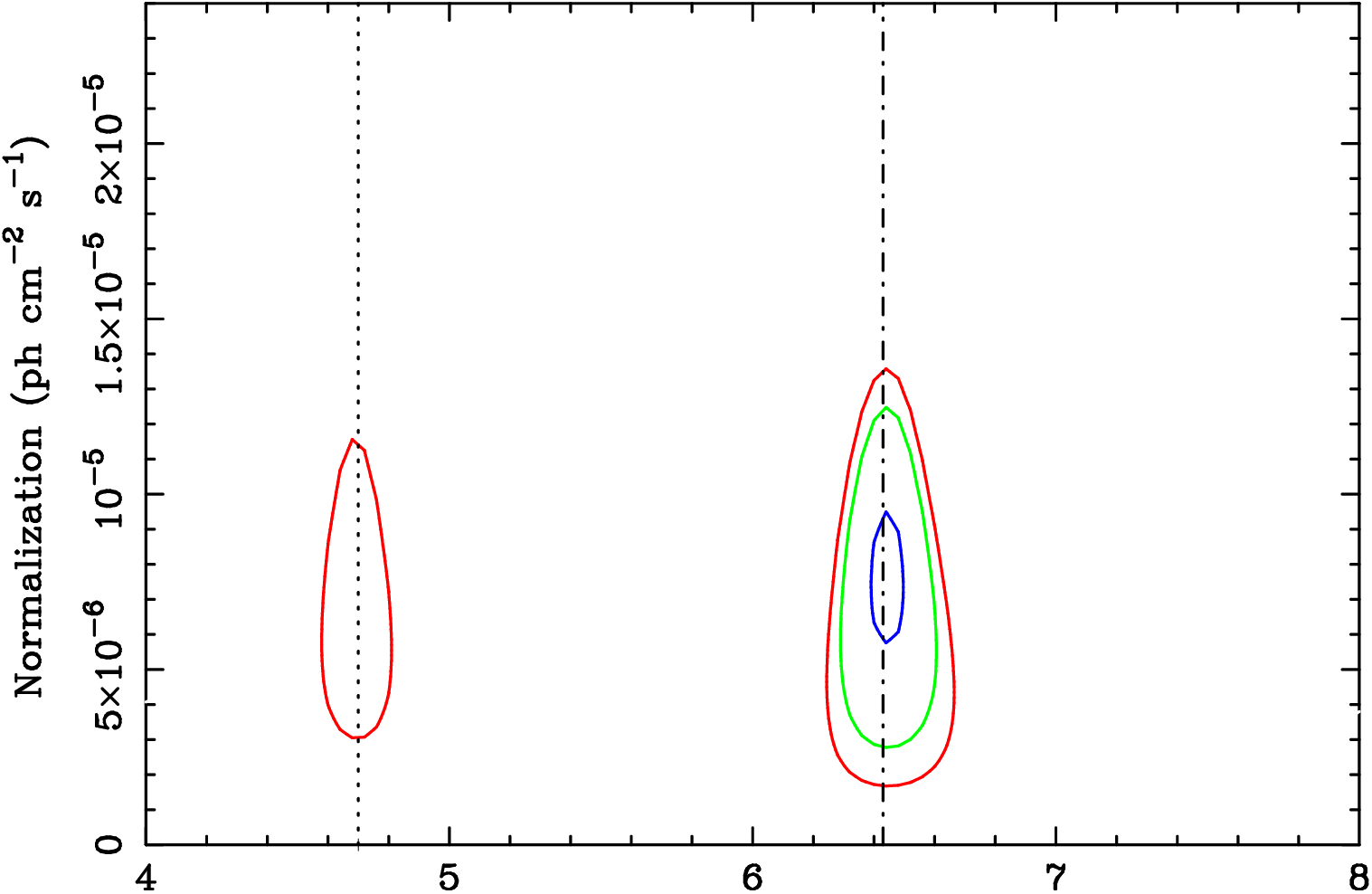,width=0.42\textwidth,height=0.26\textwidth,angle=-90}
\hspace{1.0cm}
\psfig{figure=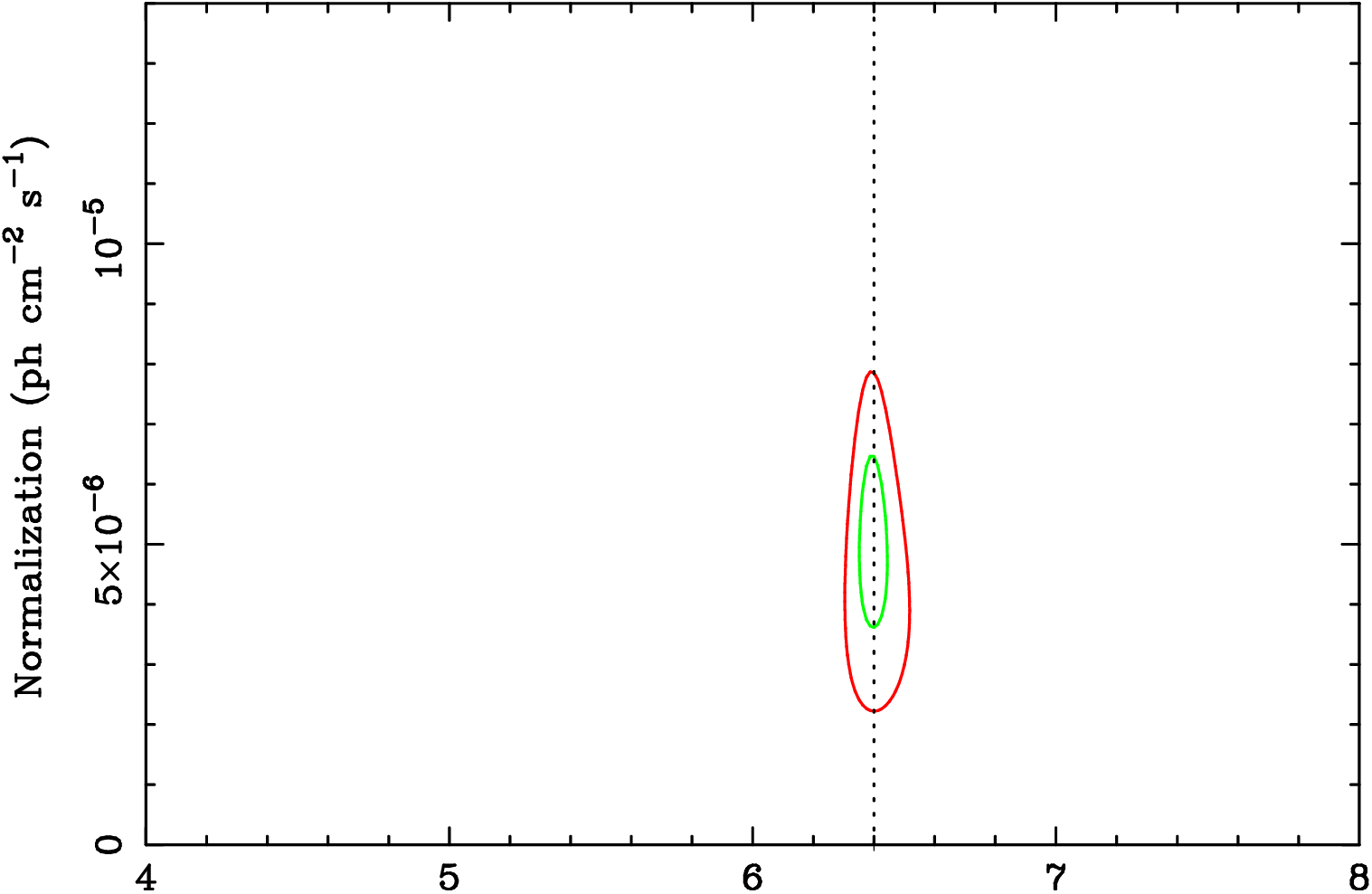,width=0.42\textwidth,height=0.26\textwidth,angle=-90} 
}}
\centerline{\mbox{ 
\psfig{figure=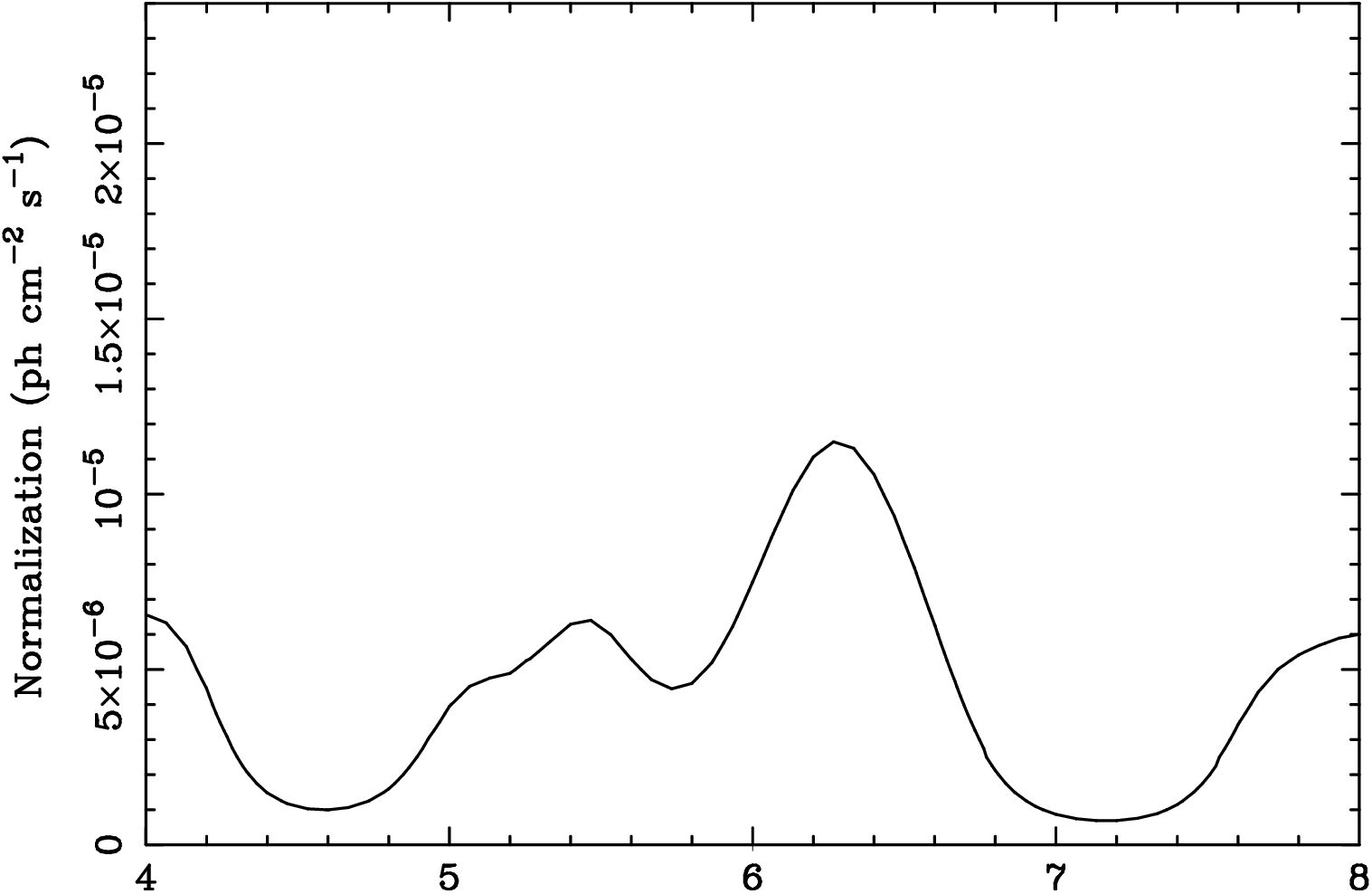,width=0.42\textwidth,height=0.26\textwidth,angle=-90}
\hspace{1.0cm}
\psfig{figure=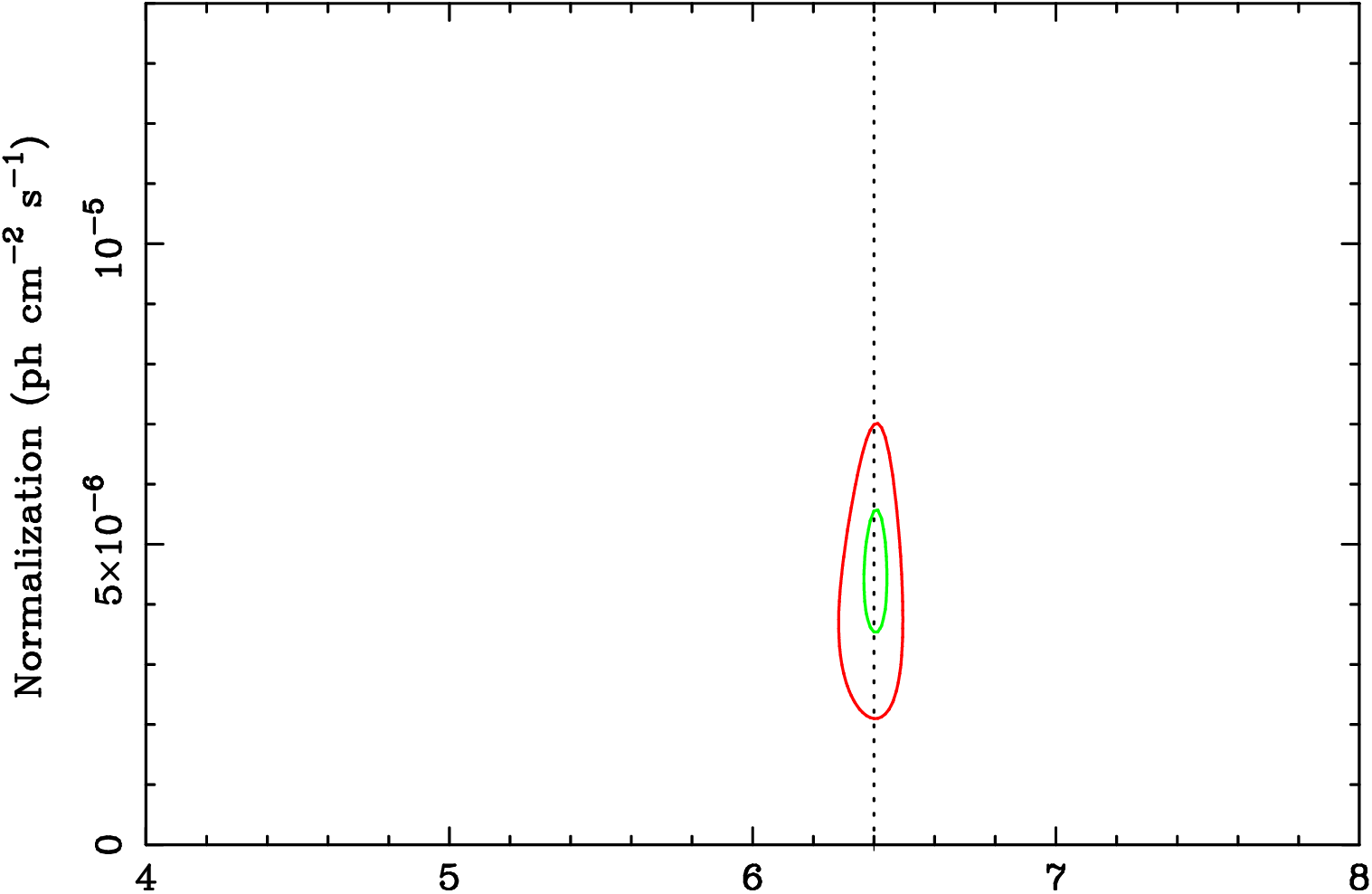,width=0.42\textwidth,height=0.26\textwidth,angle=-90} 
}}
\centerline{\mbox{ 
\psfig{figure=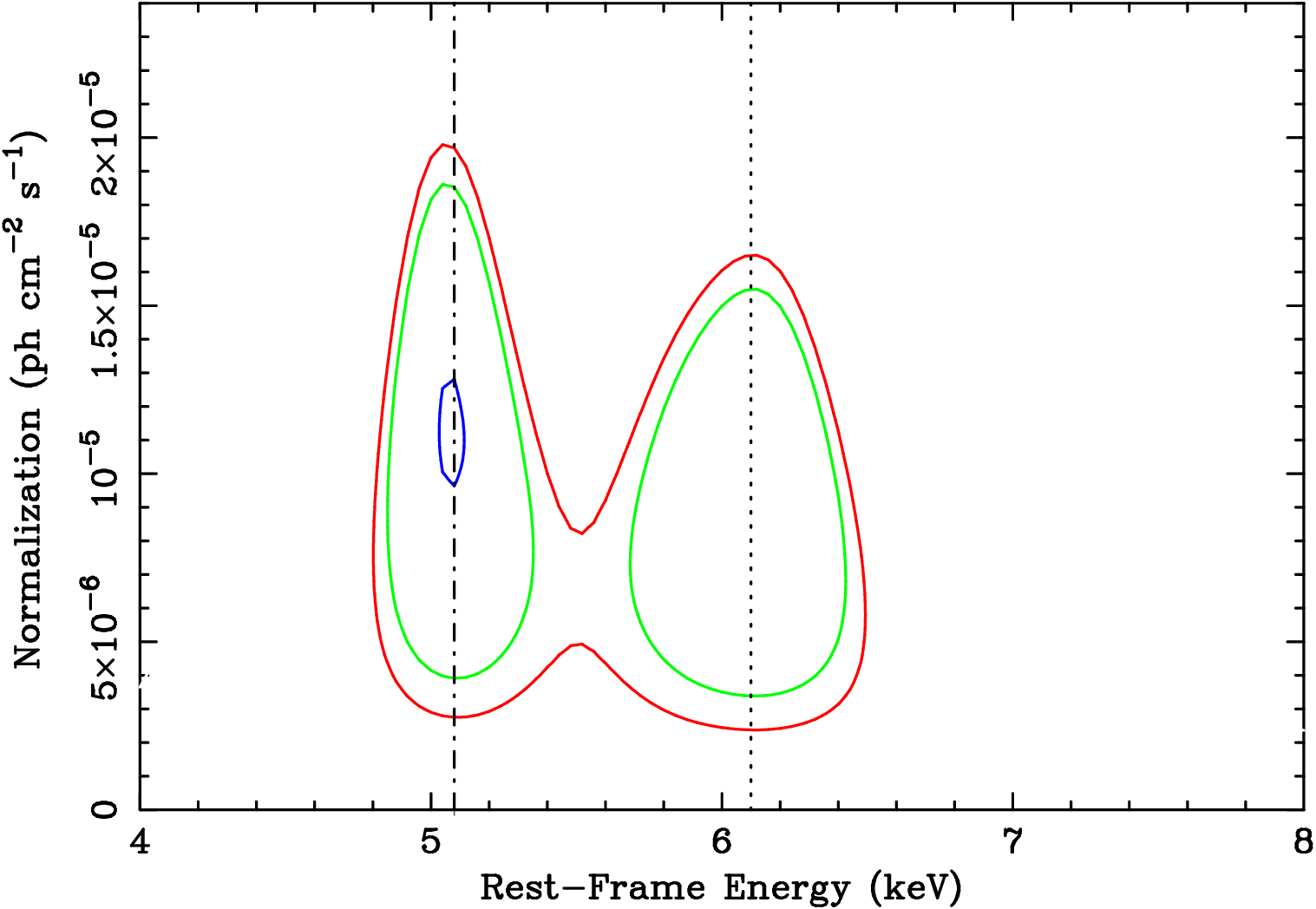,width=0.42\textwidth,height=0.27\textwidth,angle=-90}
\hspace{1.0cm}
\psfig{figure=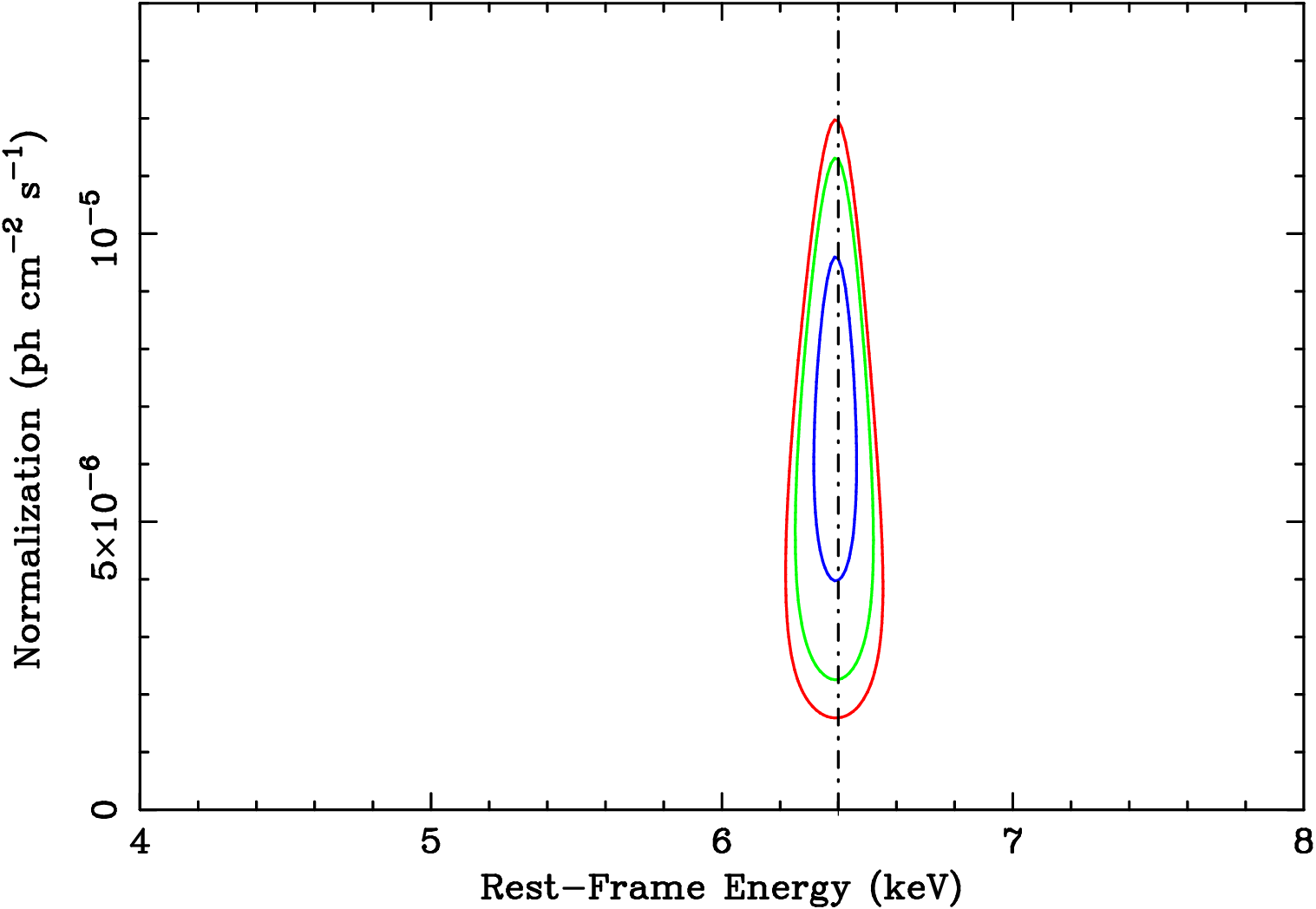,width=0.42\textwidth,height=0.27\textwidth,angle=-90} 
}}
\caption{{\bf Left Panels:} The observation of PG~1425+267 is divided
  into 4 quarters (about 12~ks total duration each but shorter net
  exposures). From top to bottom, the evolution of the iron line
  profile can be seen suggesting dramatic short--timescale
  variability. {\bf Right Panels:} Same as in the left panels, but for
  the quasar PG~1114+445, taken as comparison. The individual spectra
  have approximately the same number of counts than in the case of
  PG~1425+267 in the 4--8~keV band and are selected to be contiguous
  in the observation. Nothing like the variability suggested in
  PG~1425+267 (see left panels) can be seen.  }
\end{center}
\end{figure*}

Here we just mention that, if real, the variation would suggest that
different regions of the accretion disc are responsible for the Fe
line emission in the two intervals.  In particular, the reversal in
the prominence of the peaks (from blue to red) is a signature that, in
the second half of the observation, emission comes from a more
redshifted region of the disc than during the first half, such as a
the region on the disc receding from the observer.

\subsection{Splitting the observation in four quarters}

In order to understand if the {\emph{tentative}} variability of the
line profile is real we consider a shorter timescale analysis. It is
in fact possible that the variability happens at shorter timescales
than explored so far and that it is diluted by time--averaging on too
long intervals. We then split the observation in four equal duration
duration intervals about 12~ks long, and subtract from the four
spectra the corresponding backgrounds. Due to randomly distributed
background flares, the net exposure in the spectra is not always the
same and always shorter than 12~ks (which is the total observation
time, not the net exposure). This prevents us to investigate shorter
timescales which do not provide good enough quality spectra.

We fit the four spectra separately with the same power law plus
Galactic absorption model used for the time--averaged data and add
Gaussian emission lines if required. We also repeat the procedure that
produced Fig.~2 and present our results in the left panels of
Fig.~6. Emission lines are detected in the first, second, and fourth
interval. In the third interval, no line is clearly detected. As an indication,
we show a low--significance contour representing an improvement in
$\chi^2$ between 2.3 and 4.61 (for two degrees of freedom). 

There is no clear pattern in the evolution shown in the left panels of
Fig.~6, except that the blue peak seems to shift to lower and lower
energies as time goes on, although the energy shift is significant
only from the first ($6.6\pm 0.2$~keV) to the last quarter ($6.1\pm
0.2$~keV). The energy of the blue peak in the last quarter (lower than
6.4~keV considering the 2$\sigma$ errors) excludes that a neutral and
constant 6.4~keV Fe line from distant material contributes strongly to
the overall line profile.  Below 6~keV, the line profile variation
appears to be more erratic.  This is expected if the line comes indeed
from the inner accretion disc.  In this case, the blue peak energy is
mainly dictated by the observer inclination (the larger the
inclination the higher the maximum energy at which the line can be
seen), while the red part of the line is more sensitive to the details
of the emitting region location on the disc and is expected to vary
more erratically.

Before making any comment it is however necessary to estimate the
significance of the variations that seem to appear in the different
intervals. To do so, we make use once again of Montecarlo simulations.
The idea is to see whether the line profile is significantly different
in the four spectra. 

\subsection{Statistical significance of the variability}

We first fit jointly the four time intervals spectra with the double
Gaussian best--fit model to time--averaged data. The lines parameters
are free to vary but forced to be the same in all four intervals while
the power law slope and normalisation are allowed to be different in
the four intervals (we checked that forcing them to be the same does
not change our conclusions). In this way we are testing the goodness
of the fit under the hypothesis that the line profile is the same in
all intervals (i.e. constant in time) and we record the $\chi^2$ for
the best--fitting parameters. To test if the line profiles are
different in the four intervals (i.e. the line is variable), we then
allow the line parameters to vary independently in the four spectra,
fit again the data, and record the $\Delta\chi^2$ with respect to the
previous ``constant line profile'' fit.  We obtain $\Delta\chi^2=31.5$
which will be used as a figure for the Montecarlo simulations which
are performed as follows:

\begin{enumerate}
\item our null hypothesis is now the double Gaussian ``constant line
  profile'' best--fit model to the time--averaged spectrum. This is
  used to generate a time--averaged spectrum which is fitted to obtain
  a refined null hypothesis which includes the uncertainties due to
  photon statistics;
\item the new null hypothesis model is used to generate four simulated
  spectra with the same exposures as the four intervals selected from
  the real data and appropriate background spectra are subtracted;
\item we fit jointly the four simulated spectra forcing the line
  parameters to be the same (i.e. forcing the line profile to be
  constant in time) and record the $\chi^2$. We then allow the line
  profile to be different in the four spectra (i.e. we allow for a
  variable line profile) and record the $\Delta\chi^2$ which is
  compared to the result obtained on the real data with the same
  procedure ($\Delta\chi^2=31.5$).
\end{enumerate}
The above procedure is repeated 1000 times. In Fig.~7 we show the
$\Delta\chi^2$ distribution from the simulated spectra and compare it
with the result obtained from the data. There are 27/1000 cases in
which the resulting maximum $\Delta\chi^2$ is as good or better as the
result from the data. This means that the line profile variations in
the real data are significant at the 97.3 per cent level.

\begin{figure}
\begin{center}
 \includegraphics[width=0.27\textwidth,height=0.42\textwidth,angle=-90]{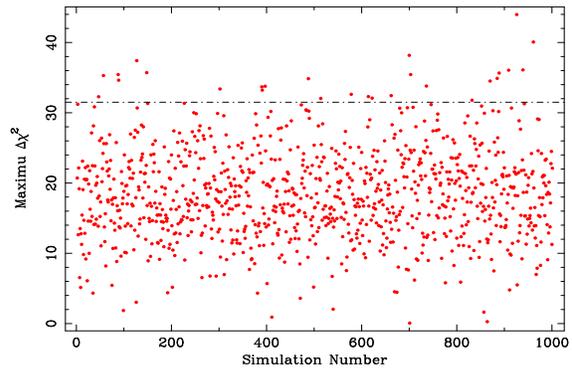}
 \caption{Distribution of the maximum $\Delta\chi^2$ obtained when the
   four simulated spectra are fitted jointly with the double Gaussian
   model but allowing the line parameters to be different in the four
   spectra. The $\Delta\chi^2$ is computed with respect to the
   best--fit double Gaussian model in which the line parameters are
   free to vary, but forced to be the same in the four spectra. The
   horizontal line is the result obtained with the same procedure on
   the real data. In 27/1000 cases we find a larger (or equal)
   $\Delta\chi^2$ than in the data, so that the line profile
   variations in the data are significant at the 97.3 per cent level.}
\end{center}
\end{figure}

\subsection{A plausibility test: the case of PG~1114+445}

Here we produce a further test for the line variability with the aim
of answering the following question: if the line profile in
PG~1425+267 was constant, would the EPIC--pn camera suggest a spurious
variability or would it be able to show that the line is indeed not
variable? This question has already been answered through Montecarlo
simulations.  However, we present also a comparison with some real
data as a complementary indication.

We searched for a 
{\it XMM--Newton} observation of a PG quasar from the sample
collected and analysed by Jim\'enez--Bail\'on et al (2005).  We were
looking for a quasar in which a narrow iron line is detected with the
same intensity (or fainter) than the 6.4~keV peak in PG~1425+267 and
in which the upper limit on the presence of a relativistic broad line
is small. This is because if the iron line is truly narrow and no or
little contribution from a broad component is seen, the most likely
origin from the line is some distant reflector such as the torus
and/or the broad line region and therefore one expects that the line
is constant in time. The line must be of the same intensity as in
PG~1425+267 to provide a fair comparison. 

The best candidate in the sample is PG~1114+445 (z=0.144). In this
quasar, Jim\'enez--Bail\'on et al (2005) report the presence of a
narrow 6.4~keV emission line with equivalent width of
$100^{+30}_{-40}$~eV, while the upper limit on a broad component is
only $33$~eV (see also Porquet et al 2004a). We downloaded the data
from the {\small{XSA}} archive and reduced them using the standard
procedures. We here report results from the EPIC--pn detector only.
The observation is about 40~ks long. We confirm the detection of a
narrow $6.40\pm 0.04$~keV line in the time--averaged spectrum with a
rest--frame equivalent width of $140^{+30}_{-60}$~eV. The line is
unresolved and the contribution from a broad component (or other
fainter peaks) in the whole 4--8~keV band is constrained to be very
small (see Fig.~8). Notice that the Fe line in PG~1114+445 has the same intensity
($4.7^{+3.1}_{-2.1}\times 10^{-6}$~ph~cm$^{-2}$~s$^{-1}$) as the
6.4~keV peak in PG~1425+267 ($5.5^{+3.6}_{-4.0}\times
10^{-6}$~ph~cm$^{-2}$~s$^{-1}$), and a slightly smaller rest--frame
equivalent width, thereby satisfying our requirements.

As for the more relevant variability study, we selected four
consecutive time--intervals in the observation. The exposure in the
individual intervals is chosen so that each spectrum has the same
number of counts in the relevant 4--8~keV band as the equivalent
spectrum obtained when the observation of PG~1425+267 is split into
quarters. In this way we make sure that the same photon statistics is
reached in the band and the quality of the data is the same. We then
repeat the same procedure applied to PG~1425+267 and report the
results in the right panels of Fig.~6.  We detect a 6.4~keV line in
all time--intervals (see Fig~6, right panels). The detection is
somewhat marginal in the first three intervals ($\Delta\chi^2 \leq 9$
for 2 more free parameters), but the presence of an emission feature
at the same energy in all intervals is, in our opinion, conclusive.

\begin{figure}
\begin{center}
 \includegraphics[width=0.27\textwidth,height=0.42\textwidth,angle=-90]{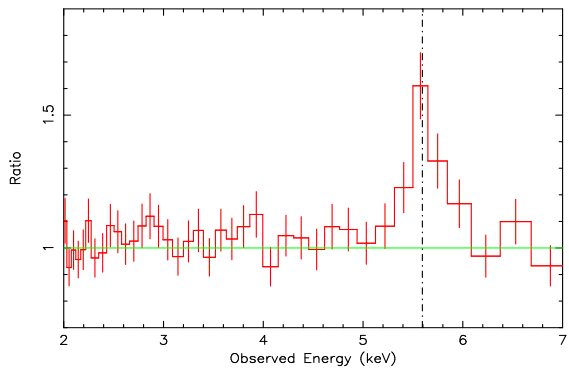}
 \caption{Ratio of the spectrum of PG~1114+445 to a simple power law
   plus Galactic absorption model. We confirm the detection of a
   narrow 6.4~keV Fe K$\alpha$ emission line. The iron K$\alpha$ line
   energy (6.4~keV) is shown in the observed frame as a vertical line.
   Data have been rebinned for visual clarity.}
\end{center}
\end{figure}

What the right panels of Fig.~6 show is that, in the case of an
emission line with the same intensity of the 6.4~keV peak in
PG~1425+267, the {\it XMM--Newton} data allow us to detect a constant
line, and to rule out any strong variability at the level seen in
PG~1425+267 (left panels of the same Figure). No shift in the line
energy is seen, contrary to the case of PG~1425+267 and no other
emission--like structures do appear in any of the intervals in the
whole 4--8~keV band. If the redshifted features seen in PG~1425+267
and the shift in the energy of the 6.4~keV peak were due to noise, it
is surprising that we do not detect any of this in the case of
PG~1114+445 given that the spectra have the same number of counts.
Thus, by comparing the left (PG~1425+267) and right (PG~1114+445)
panels of Fig.~6, it seems unlikely that the variability we are
suggesting for PG~1425+267 is completely due to noise.

Of course we can not, and do not, consider this simple test as a clear
demonstration that the line profile in PG~1425+267 is indeed rapidly
variable. However, we report here the result of this straightforward
comparison as an additional plausibility argument in favour of the
variability, the strict statistical significance of which is
established to be at the 97.3 per cent level according to our
Montecarlo simulations. Obtaining a better quality long observation of
PG~1425+267 with {\it XMM--Newton} would allow us to confirm/rule out the
short--timescale variability that is tentative and only suggested
here, and to apply more sophisticated data analysis techniques (see
Iwasawa, Miniutti \& Fabian 2004) with the potential of mapping with
great accuracy the inner accretion flow in a quasar.

\section{Some speculation}

Given that the significance of the variability is below 99 per cent, a
detailed modelling is not only difficult, but also not recommended. It
is however an interesting exercise to try to qualitatively reproduce
the apparent variations of the line profile to gain insights on the
location of the emitting region.  The tentative line profile
variations seen in Fig.~5 and 6 suggest that different regions of the
accretion disc are responsible for the bulk of the line emission at
different times.  A succession of relatively short--lived flares
illuminating different regions of the disc, one or more orbiting
reflecting spots, or disc turbulence/instabilities continuously
creating and destroying regions of enhanced emissivity at different
locations (e.g.  Armitage \& Reynolds 2003; Ballantyne, Turner \&
Young 2005) could well produce the observed line variations.  Thus, by
fitting the individual line profiles shown in the left panels of
Fig.~6 with a model accounting for non--axisymmetric emissivity on the
accretion disc, we might be able to infer some information at least on
the location of the emitting region in the different time--intervals.

We performed such an analysis by modifying our relativistic line code
(Miniutti et al 2003; Miniutti \& Fabian 2004), but the parameter
space is too large to allow for a unique solution.  We can however
report that good matches with the observed profiles are obtained by
considering emission from one (or two) spot on the disc, always
consistent with a radial distance from the black hole of about
6--15~$r_g$, but different azimuthal locations in time. It should be
noted, merely as an indication, that we never find solutions with very
localised spots on the disc, i.e. the radial and, even more
dramatically, the azimuthal extent of the emitting region is
relatively large ($1~r_g$ or more and more than 50 degrees,
respectively). The large azimuthal extent is unlikely to be due to
orbital motion because, the large mass of the black hole in this
quasar ($\sim 3\times 10^9~M_\odot$, e.g. Hao et al 2005) implies a
typical orbital timescale $\sim$~250 times longer than that we are
exploring here. It is possible that the black hole mass is not very
accurate, but it does not seem plausible that it is 100 times smaller
than estimated.

The azimuthal extent could be the signature of some elongated
structure of enhanced emissivity on the disc such as spiral waves
(Karas, Martocchia \& Subr 2001). It is also possible that the
accretion disc is not geometrically thin (or not completely) and that
some elevated regions intercept more efficiently than their
surroundings the X--ray illuminating flux coming from the centre. The
effect may be strong if the primary X--ray emission comes from very
close to the central black hole because, in this case, much of the
radiation is beamed along the equatorial plane, so that even small
elevations on the disc would make a dramatic difference in the
reflection spectrum (see e.g. Miniutti \& Fabian 2004). 

If elongation due to orbital motion is excluded (see above), and if
the variability is associated with changes in the illumination of the
disc by flares above it, the extent of the emitting regions can be
used to argue against a very small height of the flares above the
disc.  A flare located at very small height would produce a much
smaller spot than observed.  This result is in line with what was
found in the case of NGC~3516, in which the iron line variability was
best explained by a large spot such as that produced by a corotating
flare at few $r_g$ above the accretion disc (Iwasawa, Miniutti \&
Fabian 2004).

On the other hand, one large spot can be successfully reproduced by
considering a large number of small height flares occurring close to
each other in one main region. Notice that a large number of
independent flares would violate the log--normal distribution of
fluxes that is inferred for both AGN and Galactic BHC from the
observed rms--flux relation (Uttley, McHardy 2001; Uttley, McHardy \&
Vaughan 2005).  Indeed, the log--normal distribution strongly argues
against the presence of more than 2--3 independent flares. Therefore,
if we persist to interpret the emitting regions as due to flares above
the disc (rather than structures on the disc itself, independently of
the illumination), we can retain only two possibilities: i) one (or
two) independent flares are located at a relatively large height (few
gravitational radii) above the disc, or ii) a large number of
localised, small--height flares occur close to each other in one (or
two) main independent regions, but the individual flares in each
region are not independent.

It is interesting to note that the ``thundercloud model'' proposed by
Merloni \& Fabian (2001) seems to lie in between the two cases. In
this model, magnetic flares produce the X--ray continuum via inverse
Compton scattering of soft seed photons. The difference with other
similar models is that the fundamental heating event must be compact,
with size comparable to the disc thickness, but height at least one
order of magnitude larger.  Moreover, the heating event (most likely
due to magnetic reconnection) is shown to proceed in correlated trains
(such as in an avalanche) with the size of the avalanche determining
the size and luminosity of the overall active region. If the number of
simultaneous independent active regions is limited (and only one or
two regions are required by the data) the model seems able to
reproduce the variability we are suggesting here, potentially
preserving at the same time the rms--flux relation and log--normal
distribution of fluxes.


\section{The broadband spectrum and warm absorber of PG~1425+267}

\begin{figure}
\begin{center}
 \includegraphics[width=0.27\textwidth,height=0.42\textwidth,angle=-90]{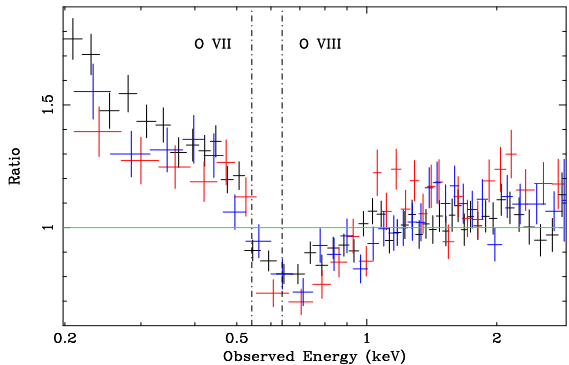}
 \caption{Data to model ratio for the three EPIC detectors when the
   hard pn model is extrapolated in the soft band. The deficit above
 0.5~keV (0.683~keV in the rest--frame) is the signature of absorption
 by warm photoionized gas. We also show as vertical lines the
 energies of the expected oxygen edges in the observed frame.}
\end{center}
\end{figure}

To better constrain the soft spectrum of PG~1425+267, we extracted the
MOS spectra as well.  If the model we used so far is extended to the
soft band, a clear deficit remains in the 0.5--1~keV band (observed
frame), most likely due to intervening absorption in a warm absorber
(see Fig.~5 in which the extension to the soft spectrum is shown for
all {\it XMM--Newton} X--ray detectors). Extending the pn data below
0.5~keV is however not recommended because of some residual
calibration problems that still affect the pn camera below that
energy. Therefore, we used the pn data in the 0.5--8.5~keV range only,
while the MOS data are used in the 0.2--8.5~keV (observed energy
ranges). With the addition of one absorption edge, the fit is
acceptable ($\chi^2 = 1044$ for 939 degrees of freedom).  However, the
energy of the edge is $0.79 \pm 0.03$~keV, which is not consistent
with either O~{\footnotesize{VII}} ($0.74$~keV) nor
O~{\footnotesize{VIII}} ($0.87$~keV), and the depth is very large
($\tau=0.8 \pm 0.1$). This probably indicates the simultaneous
presence of two edges at energies below and above that measured. We
then add a second edge, and fix both energies at the rest--frame
values for O~{\footnotesize{VII}} and O~{\footnotesize{VIII}}. With
this two--edges model, we find an improvement of $\Delta\chi^2=16$ for
the same number of degrees of freedom with respect to the single--edge
model. The depth of the edges are $\tau_{\rm OVII} = 0.5 \pm 0.2$ and
$\tau_{\rm OVIII} = 0.4 \pm 0.2$. If the energy of the edges are free
to vary, they tend to slightly higher values, but are still consistent
with the rest--frame energies within the errors.  Very little soft
excess is present in the MOS data. However, the addition of a
(redshifted) blackbody component results in a further improvement in
the fit (at more than the 99 per cent level according to the F--test).
The temperature of the thermal emission component is $80\pm 10$~eV,
cooler than the typical temperature found by Porquet et al (2004a) and
Piconcelli et al (2005) in their analysis of the {\it XMM--Newton}
sample of PG quasars (about $150$~eV). The final power law slope is
$\Gamma=1.55 \pm 0.03$ and the overall fit is excellent with $\chi^2 =
1004$ for 937 degrees of freedom.

\section{Conclusions}

The hard spectrum of PG~1425+267 is characterised by the presence of
an emission feature which is best described by a double--peaked Fe
line profile. The blue peak of the line is consistent with an energy
of 6.4~keV and an unresolved width, and its origin could well be in
some distant reflector. The red peak of the line (at 5.3~keV) is
however redshifted and broad therefore suggesting emission from the
accretion disc.  According to Montecarlo simulations, the broad and
redshifted part of the line profile is significant at the 99.1 per
cent level. The simplest interpretation of the time--averaged data is
that the accretion disc is acting as a reflector producing an iron
K$\alpha$ line whose profile is dictated by Doppler and gravitational
shifts in the orbiting gas. The best--fitting model we find is that of
a relativistic line from an accretion disc around a Schwarzschild
black hole providing a statistical improvement with respect to a
simple power law model which is significant at more than the 99.99 per
cent level (F--test). No black hole spin is required by the data,
although emission from within the marginal stable orbit of a
non--rotating black hole cannot be excluded. Due to the quality of the
data, it is not possible to exclude a partial covering model plus
narrow 6.4~keV line as an alternative to the relativistic line. However, the
partial covering model is largely unconstrained and is not able to
well reproduce the broad emission feature at 5.3~keV, suggesting that
the relativistic line is a better explanation of the hard spectrum.
The detection of a broad relativistic line in a radio--loud quasar
argues against a scenario in which, due e.g.  to low mass accretion
rate, the disc in such objects is truncated (see also the discussion
in Ballantyne \& Fabian 2005).

When the observation is split into quarters, the energy of the blue
peak appears to shift from $6.6 \pm 0.2$~keV (first quarter of the
observation) to $6.1 \pm 0.2$~keV (last quarter) in less than 48~ks.
The fast variability and the energy of the blue peak in the last
quarter excludes a strong contribution to the line profile from a
constant 6.4~keV iron line from distant material. The large majority
(if not all) of the line profile is therefore likely emitted from a
highly dynamical medium such as the accretion disc. Moreover bright
redshifted peaks appear at $4.7\pm 0.2$~keV and, with higher
significance, at $5.1^{+0.1}_{-0.2}$~keV in the second and last
quarter of the observation respectively. We performed simulations to
establish the significance of the apparent variations in the line
profile and found that they are significant at the 97.3 per cent
level.

If the variability is real, it indicates
emission from different regions on the accretion disc at different
times in a range of 6--15~$r_g$ from the central black hole. Our
understanding of the inner accretion flow in PG~1425+267 would benefit
very much from a future, long, and better quality {\it XMM--Newton}
observation that would allow us to secure (or rule out) any
variability on firm statistical grounds. It is clear (also from
previous recent results, see Iwasawa, Miniutti \& Fabian 2004 for the
remarkable case of NGC~3516) that the potential of future missions
with much larger collecting area than {\it XMM--Newton} in the Fe K
band, such as XEUS and Constellation--X, is outstanding. The prospects
of probing the strong gravity regime of General Relativity via X--ray
observations of relativistic Fe lines and of their short timescale
variability look stronger now than ever before.

\section*{Acknowledgements}
Based on observations obtained with {\it XMM--Newton}, an ESA science
mission with instruments and contributions directly funded by ESA
Member States and NASA. GM thanks the PPARC and ACF the Royal Society
for support. We thank the anonymous referee for many suggestions that
improved our paper.

\end{document}